\documentclass[aps,prc,preprint,groupedaddress]{revtex4}

\usepackage{graphicx}

\newcommand{\al}{\alpha}

\newcommand{\oeq}{\begin{equation}}
\newcommand{\ceq}{\end{equation}}
\newcommand{\oeqn}{\begin{eqnarray}}
\newcommand{\ceqn}{\end{eqnarray}}

\renewcommand{\>}{\rangle}
\newcommand{\<}{\langle}
\renewcommand{\(}{\left(}
\renewcommand{\)}{\right)}
\renewcommand{\[}{\left[}
\renewcommand{\]}{\right]}

\newcommand{\stf}{\,\,\,}

\newcommand{\stb}{\!\!\!}

\newcommand{\kal}{|\al\>}

\newcommand{\knu}{|\nu\>}

\newcommand{\bmu}{\<\mu|}
\newcommand{\bnu}{\<\nu|}

\newcommand{\oQ}{\hat{Q}}

\newcommand{\oH}{\hat{H}}

\newcommand{\ox}{\hat{x}}
\newcommand{\oy}{\hat{y}}
\newcommand{\oz}{\hat{z}}

\newcommand{\oV}{\hat{V}}

\newcommand{\oZ}{\hat{Z}}

\newcommand{\ovr}{\hat{\bf r}}

\renewcommand{\d}{{\mbox d}}

\newcommand{\hb}{\hbar}

\renewcommand{\vr}{{\bf r}}

\begin{document}

\title{Couplings between dipole and quadrupole vibrations in tin isotopes}

\author{C. Simenel} 
\affiliation{CEA, Centre de Saclay, IRFU/Service de Physique Nucl\'eaire, F-91191 Gif-sur-Yvette, France.}
\author{Ph. Chomaz}
\affiliation{CEA, Centre de Saclay, IRFU/Dir, F-91191 Gif-sur-Yvette, France.}
\affiliation{GANIL (DSM-CEA/IN2P3-CNRS), B.P. 55027, 
F-14076 Caen cedex 5, France}

\date{\today}

\begin{abstract}
We study the couplings between collective vibrations such as the isovector giant dipole and isoscalar giant quadrupole 
resonances in tin isotopes in the framework of the time-dependent Hartree-Fock theory with a Skyrme 
energy density functional.
These couplings are a source of anharmonicity in the multiphonon spectrum. 
In particular, the residual interaction is known to couple the isovector giant dipole resonance with the isoscalar giant quadrupole resonance built on top of it, 
inducing a nonlinear evolution of the quadrupole moment after a dipole boost.
This coupling also affects the dipole motion in a nucleus with a static 
or dynamical deformation induced by a quadrupole constraint or boost respectively. 
Three methods associated with these different manifestations of the coupling
are proposed to extract the corresponding matrix elements of the residual interaction. 
Numerical applications of the different methods to $^{132}$Sn are in good agreement with each other.
Finally, several tin isotopes are considered to investigate the role of isospin and mass number on this coupling.
A simple $1/A$ dependence of the residual matrix elements is found with no noticeable contribution from the isospin. 
This result is interpreted within the Goldhaber-Teller model.
\end{abstract}


\maketitle

\section{Introduction}

A particular interest in strongly interacting systems is their ability to present disorder or chaos, 
and, in the same excitation energy range, well-organized motion.
Atomic nuclei are known to show both behaviors~\cite{boh75}.
In particular, they exhibit a large variety of collective vibrations,
also called giant resonances (GRs),
with excitation energy usually above the particle emission threshold~\cite{har01}.
The GRs are associated 
with anomalously large cross sections in some nuclear reactions. 

Baldwin and Klaiber observed the isovector giant dipole resonance (GDR) 
in photofission of uranium nuclei~\cite{bal47}, 
interpreted as a vibration of neutrons against protons~\cite{gol48}. 
This GDR was investigated with several probes~\cite{har01}
and was also observed on top of highly excited states, e.g., in hot nuclei~\cite{new81}.
The survival of ordered motion in hot nuclei, i.e., in a chaotic environment, 
is one of the most striking phenomena in nuclear physics.
Other kinds of GR have been discovered, such as the isoscalar giant quadrupole resonance (GQR) 
associated with an oscillation of the shape between a prolate and an oblate deformation~\cite{fuk72},
and the isoscalar giant monopole resonance (GMR) corresponding to a breathing mode~\cite{mar76,har77,you77}.

The GRs are usually associated with the first phonon of a small-amplitude harmonic motion.
However, the proof of their vibrational nature came with the observation 
of their two- and three-phonon states~\cite{cho95,aum98,sca04}.
Multiphonon studies also provided a good test of the harmonic picture. 
In particular, anharmonicity was found in an abnormally large excitation probability of these states, 
indicating that different phonon states couple because of the residual interaction~\cite{vol95,bor97}. 
Microscopic investigations, such as the random phase approximation (RPA)
together with boson mapping techniques~\cite{fal03} and the nonlinear response
to an external field in the time-dependent Hartree-Fock (TDHF) theory~\cite{sim03,cho04} 
showed, indeed, that strong couplings between GMR, GQR and GDR occur.
In particular, a GMR or a GQR (resp. a GMR) can be excited on top of a GDR (resp. a GQR),
leading to couplings between one- and two-phonon states.
As a consequence, GRs cannot be described in a purely harmonic picture. 
Anharmonicities were also found to affect pygmy dipole resonances,
though depending on the choice of the nuclear functional \cite{lan09}.

The goal of the present work is to get a deeper insight into the couplings between various GRs,
which represents a first step toward understanding complexity and disorder in nuclei at high excitation energies.
As an example, we focus on the coupling between isovector dipole and isoscalar quadrupole vibrations. 
A clear link between the linear dipole motion on a deformed state and
the quadratic response of the quadrupole moment to an external dipole excitation
(investigated in Ref.~\cite{sim03}) is made. 
The TDHF theory is used to compute the residual interaction
coupling the one-phonon state of the GDR to the two-phonon state with a GQR built on top of the GDR. 
Applications to spherical tin isotopes are performed 
to investigate the role of the isospin degree of freedom 
and of the total number of nucleons on the coupling.

We present a schematic model describing couplings between GRs
and their effect on one-body observables in Sec.~\ref{sec:model}.
The TDHF formalism and its application to nuclear vibrations are discussed in Sec.~\ref{sec:tdhf}.
Numerical details on the 3-dimensional TDHF code are also given.
A detailed investigation of the couplings in $^{132}$Sn is presented in Sec.~\ref{sec:results}, 
together with a more systematic analysis in tin isotopes.
Finally, we conclude in Sec.~\ref{sec:conclusion}.

\section{A schematic model for GR coupling}
\label{sec:model}


Let us illustrate the effect of couplings between vibrational modes within a simple schematic model introduced in Ref.~\cite{sim03}.
We consider the Hamiltonian 
\oeq 
\oH=\oH_0+\oV
\label{eq:H}
\ceq
where $\oH_0$ corresponds to the harmonic (HF+RPA) part 
and the residual interaction $\oV$ couples collective modes.
Eigenstates of $\oH_0$ are one- and two-phonon states $\knu$ and $|\nu\mu\>$ with eigenenergies
$E_\nu=E_0+\hb\omega_\nu$ and $E_{\nu\mu}=E_0+\hb\omega_\nu+\hb\omega_\mu$, respectively,
where $\omega_{\mu,\nu}$ denote the collective frequencies and $E_0$ is the ground-state energy. 
In the following, $\hb$ is omitted in the notation.
Only the coupling between the two states $\knu$ and $|\nu\mu\>$ is considered here. 
The associated matrix element of the residual interaction is noted $v_\mu=\bnu \oV |\nu \mu\>$.
Such couplings between one- and two-phonon states have
been proven to be the most important one in nuclei~\cite{fal03}.
Using first-order perturbation theory, 
the eigenvalues of $\oH$ are those of $\oH_0$ with eigenstates
\oeq 
|\overline{\nu}\> \approx \knu -\varepsilon_\mu |\nu\mu\>
\ceq
and 
\oeq 
|\overline{\nu\mu}\> \approx |\nu\mu\> + \varepsilon_\mu |\nu\>
\ceq
where $\varepsilon_\mu=\frac{v_\mu}{\omega_\mu}$.


The couplings are expected to affect the evolutions of expectation values 
of one-body observables such as the multipole moments $Q_\nu(t)\equiv\<\oQ_\nu\>(t)$.
We investigate below three different manifestations of the couplings on these evolutions. 
They will be used in the next section to compute~$v_\mu$ from TDHF calculations
in the case of coupling between giant dipole and quadrupole resonances. 

\subsection{Quadratic response}
\label{subsubsec:quadratic}

The effect of couplings in the quadratic response has been introduced in Ref.~\cite{sim03}. 
Highlights on the main steps are given here.
At initial time, the ground state~$|0\>$ of the system is excited by a boost 
with the one-body operator~$\oQ_\nu$
\oeq
|\Psi(0)\> = \exp(-ik_\nu\oQ_\nu) |0\>.
\label{eq:boost}
\ceq 
Developing the exponential up to second order in the boost intensity~$k_\nu$
and considering an evolution under the Hamiltonian defined in Eq.~(\ref{eq:H}), 
the state at time $t$ reads at first order in $\varepsilon_\mu$
\oeq
|\Psi(t)\> \approx \exp(-iE_0t) \[\(1-\frac{k_\nu^2q_\nu^2}{2}\) |0\> 
-ik_\nu q_\nu e^{-i\omega_\nu t} \(|\overline{\nu}\> -\varepsilon_\mu e^{-i\omega_\mu t} |\overline{\nu\mu}\>\)\],
\ceq
where $q_\nu=\bnu\oQ_\nu|0\>$ is the transition amplitude that we assume to be real.

The expectation value of the one-body observable used in the boost exhibits oscillations. 
Indeed, in case of no static deformation in the ground state, we have
\oeq
Q_\nu(t) =-2 k_\nu q_\nu^2 \sin(\omega_\nu t) + O(k_\nu^3).
\label{eq:linear}
\ceq
In particular, its amplitude increases linearly with the boost intensity in the small amplitude regime.
In addition to this linear response, the coupling induces an oscillation of the other collective mode $Q_\mu$:
\oeq
Q_\mu(t) \approx 2 k_\nu^2 q_\nu^2 q_\mu \frac{v_\mu}{\omega_\mu} \[\cos(\omega_\mu t)-1\]
\label{eq:quadratic}
\ceq
where we have assumed 
$q_\mu=\bmu\oQ_\mu|0\> =\<\mu\nu|\oQ_\mu\knu$.
This oscillation is then quadratic in~$k_\nu$ and provides a first method to compute the residual interaction~$v_\mu$,
assuming the fact that a nonlinear theory, such as TDHF, 
is used to follow the expectation values of the one-body observables.  
We finally note that $Q_\nu(t)$ and $Q_\mu(t)$ have different frequencies and start in phase quadrature.

\subsection{Linear response in an external static field}
\label{subsubsec:deformed}

It is interesting to note that the coupling may also manifest itself in the linear 
response to the boost~(\ref{eq:boost}) if an external static field is added to the Hamiltonian~(\ref{eq:H})
\oeq
\oH(\lambda) = \oH(0) + \lambda \oQ_\mu.
\ceq
We choose $\lambda$ small enough to induce a linear static deformation defined as
\oeq
Q_\mu^0(\lambda) = \<0(\lambda)|\oQ_\mu|0(\lambda)\> \approx \lambda \(\frac{\partial Q_\mu^0}{\partial \lambda}\)_{\lambda=0},
\ceq
where the ground state $|0(\lambda)\>$ of $\oH(\lambda)$ contains a contribution 
of the one-phonon state $|\mu\>$:
\oeq
|0(\lambda)\> \approx |0\> + \frac{\lambda}{2q_\mu}\(\frac{\partial Q_\mu^0}{\partial \lambda}\)_{\lambda=0}|\mu\>.
\ceq
The external potential modifies linearly the eigenenergies of the Hamiltonian
and the frequency of the linear response to a boost~(\ref{eq:boost}) on $|0(\lambda)\>$ 
follows
\oeq
\(\frac{\partial \omega_\nu}{\partial \lambda}\)_{\lambda=0} = \frac{v_\mu}{q_\mu}\(\frac{\partial Q_\mu^0}{\partial \lambda}\)_{\lambda=0},
\label{eq:slope}
\ceq
providing another direct way to extract the matrix element $v_\mu$ of the residual interaction. 
We emphasize the fact that, here, the nonlinear response is not invoked and a RPA code
allowing static deformation in the ground state would be sufficient to compute such couplings.

\subsection{Response to two simultaneous excitations}
\label{subsubsec:double_boost}

We showed two manifestations of the coupling $(i)$ in the quadratic response and $(ii)$ in the linear response under a static constraint.
Let us now introduce a third one where the response $Q_\nu(t)$ is studied after a simultaneous double boost of~$\oQ_\mu$ and~$\oQ_\nu$: 
\oeq
|\Psi(0)\> = e^{-ik_\mu\oQ_\mu}e^{-ik_\nu\oQ_\nu} |0\>.
\label{eq:double_boost}
\ceq
The $\oQ_\mu$ term modifies the response of Eq.~(\ref{eq:linear}) with an additional term
\oeqn
\Delta Q_\nu(t)&=&\<\oQ_\nu\>(t)-\<\oQ_\nu\>_{k_\mu=0}(t) \nonumber \\
&=& 4k_\nu k_\mu q_\nu^2 q_\mu \frac{v_\mu}{\omega_\mu} \[1-\cos(\omega_\mu t)\]\cos(\omega_\nu t).
\ceqn
It is convenient to write this evolution with the form
\oeqn
x(t)&=&\frac{Q_\nu(t)}{\overline{Q}_{\nu}}\nonumber \\
&=&\sin\omega_\nu t-\beta \cos\omega_\nu t+\frac{\beta}{2}\cos(\omega_\mu+\omega_\nu)t +\frac{\beta}{2}\cos(\omega_\mu-\omega_\nu)t 
\label{eq:x}
\ceqn
where $\overline{Q}_{\nu}=-2k_\nu q_\nu^2$ and $\beta=2k_\mu q_\mu v_\mu/\omega_\mu$.
In fact, we can show that $x(t)$ is a solution of the differential equation 
\oeq
\frac{\ddot{x}}{\omega_\nu^2}+\[1-2\beta\frac{\omega_\mu}{\omega_\nu}\sin \omega_\mu t\]x+\beta \frac{\omega_\mu^2}{\omega_\nu^3}\dot{x} \cos \omega_\mu t=0
\ceq
if one keeps only the first-order terms in~$\beta$.
The two first terms of the left-hand side are equivalent to a Mathieu's equation.
It is not surprising because the latter has been shown to qualitatively reproduce 
the preequilibrium dipole motion coupled to collective shape vibrations 
of the system in $N/Z$ asymmetric fusions~\cite{sim01,sim07}. 

We see in Eq.~(\ref{eq:x}) that the effect of the coupling produces 
vibrations at frequencies $|\omega_\nu\pm\omega_\mu|$. 
By analogy to the standard response function related to the strength distribution~\cite{rin80}
we introduce the coupling response function
\oeq
R^c_\nu(\omega) = \frac{-1}{\pi k_\nu k_\mu} \int_0^\infty \stb \d t \stf \cos (\omega t) \Delta Q_\nu(t)
\label{eq:coupling}
\ceq
defined for $\omega\ge 0$. The latter can be used to investigate the coupling because it is linearly proportional to $v_\mu$:
\oeq
R^c_\nu(\omega) \approx \frac{q_\nu^2 q_\mu v_\mu}{\omega_\mu}\[-2 \delta(\omega_\nu-\omega)+\delta(\omega_\nu+\omega_\mu-\omega)+\delta(|\omega_\nu-\omega_\mu|-\omega)\].
\label{eq:resp_func}
\ceq
Equation~(\ref{eq:resp_func}) provides then a third way to extract~$v_\mu$.

Let us finally note that the contribution to the coupling response function at $\omega_\nu$ and those at $|\omega_\nu\pm\omega_\mu|$  have opposite signs in Eq.~(\ref{eq:resp_func}) and that the integral of the coupling response function is zero. 
It is interesting to note that this property is still valid at all order in $k_\nu$ and $k_\mu$.
To show it, let us recall that, in our schematic model, $\oQ_\nu|0\>=q_\nu\knu$ and $\oQ_\nu^2|0\>=q_\nu^2|0\>$, 
which implies $e^{-ik_\nu\oQ_\nu}|0\>= c_\nu |0\> -i s_\nu \knu$, where $c_\nu=\cos (k_\nu q_\nu)$ and $s_\nu=\sin (k_\nu q_\nu)$.
The response $Q_\nu(t)$ following the double boost of Eq.~(\ref{eq:double_boost}) then becomes
\oeq
Q_\nu(t) = -2c_\nu s_\nu q_\nu \sin(\omega_\nu t) + 4c_\nu c_\mu s_\nu s_\mu q_\nu \frac{v_\mu}{\omega_\mu} \[1-\cos(\omega_\mu t)\]\cos(\omega_\nu t).
\ceq
We see that Eq.~(\ref{eq:x}) is still valid if one replaces 
$\overline{Q}_{\nu}$ by $\overline{Q}_{\nu}'=-2c_\nu s_\nu q_\nu$ and 
$\beta$ by $\beta'=2c_\mu s_\mu v_\mu/\omega_\mu$.
Then, the $\omega$-dependance of Eq.~(\ref{eq:resp_func}) is  unchanged.
As a consequence, the cancellation of the integral of the coupling response function defined in Eq.~(\ref{eq:coupling}) is not limited to the small-amplitude regime. 

\section{the time-dependent Hartree-Fock approach}
\label{sec:tdhf}

\subsection{Applications to nuclear vibrations}

Coherent motion of fermions such as collective vibrations in nuclei
can be modeled by time-dependent mean-field approaches like the 
TDHF theory  proposed by Dirac~\cite{dir30}. 
Indeed, in its linearized version, TDHF is equivalent to the
RPA which is the basic tool to understand the
collective vibrations in terms of independent phonons.

As we saw in the previous section, giant resonance properties can be investigated
by studying the response of the system to 
an external (collective) one-body field. In particular, time evolution 
of one-body (collective) observables, which can be computed using mean-field approximations,
contain the necessary information to investigate the couplings between collective modes. 
Indeed, TDHF takes into account the effects of the residual
interaction if the considered phenomenon can be observed in the time
evolution of a one-body observable. In particular, 
the nonlinear response in TDHF contains the couplings
between one- and two-phonon states coming from the 3-particle 1-hole and
1-particle 3-hole residual interaction~\cite{sim03}.
In that sense, it goes beyond the RPA, which is a harmonic picture 
and contains only 1-particle 1-hole residual interaction. 

In its unrestricted form (i.e., with no constraint on spatial symmetry), 
TDHF authorizes all possible spatial forms of the nucleon wave functions,
which is crucial because of both shell the effects and the wave dynamics. 
In addition, Landau spreading and evaporation damping are well accounted for~\cite{cho87}. 
However, it does not incorporate the dissipation from two-body mechanisms~\cite{gon90,won78,lac98}. 
Inclusion of pairing correlations is possible within the time-dependent 
Hartree-Fock-Bogolyubov theory~\cite{ave08}, but realistic applications in three 
dimensions are not yet achieved.
Extension to theories going beyond the one-body limit such as extended
TDHF~\cite{lac98}, second RPA~\cite{dro90,lac00}, time-dependent density matrix theory~\cite{toh01,toh02,wan85}
or stochastic one-body transport theory~\cite{lac01} should be considered for
realistic description of giant resonance properties~\cite{lac07}.
 
Application of TDHF to nuclear dynamics has been possible thanks to the Skyrme-type 
effective interaction~\cite{eng75,bon76}.
Early realistic TDHF codes have been applied 
to study collective vibrations in nuclei with simplified Skyrme interactions~\cite{blo79}.
Recent increase of computational power allowed 
realistic TDHF description of  giant resonances in 3 dimensions 
with full Skyrme energy density functional 
(EDF)~\cite{sim03,uma05,mar05,nak05,sim08b}.
In particular, TDHF has been used to investigate nonlinear effects in nuclear vibrations~\cite{sim03,rei07}.

\subsection{Formalism}

The TDHF equation can be written as a Liouville-Von Neumann equation 
\begin{equation}
i \frac{\partial}{\partial t} \rho = \left[h[\rho],\rho\right],
\label{eq:tdhf}
\end{equation}
where $\rho$ is the one-body density matrix of an independent particles state with elements 
\begin{equation}
\rho(\mathbf{r} sq, \mathbf{r'}s'q') = \sum_{i=1}^{A} \varphi_i(\mathbf{r} sq)\varphi_i^*(\mathbf{r'}s'q'),
\end{equation}
where $A$ is the number of nucleons.
The sum runs over all occupied single-particle wave functions $\varphi_i$ and $\mathbf{r}$, $s$, and $q$ denote the nucleon position, spin, and isospin respectively.
The Hartree-Fock single-particle Hamiltonian $h[\rho]$ is related to the EDF, denoted by $E[\rho]$, through
\begin{equation}
h[\rho](\mathbf{r} sq, \mathbf{r'}s'q') = \frac{\delta E[\rho]}{\delta \rho(\mathbf{r'} s'q', \mathbf{r} sq)}.
\end{equation}

\subsection{Numerical details}

In this work, the TDHF equation~(\ref{eq:tdhf}) is solved iteratively in time on a spatial grid with a plane of symmetry using the {\textsc{tdhf3d}} code built by P.~Bonche and coworkers~\cite{kim97} with the SLy4 parametrization of the Skyrme EDF~\cite{cha98}.
The latter has been constrained on the pure neutron matter equation of state to improve the description of exotic nuclei. For instance, together with a density-dependent zero-range pairing interaction, it allows a somewhat good reproduction of the isotopic shifts between proton and neutron mean-square radii in lead isotopes. It also improves the description of the isotopic evolution of the binding energies~\cite{cha98}. Defining the neutron drip line as the isotope for which the chemical potential vanishes, it is estimated to be around $^{176}$Sn in the tin isotopic chain~\cite{ben03}, though it might depend on the choice of the pairing interaction and on the inclusion of beyond mean-field correlations.

Good convergences of the quadrupole and dipole moment evolution is ensured with a lattice spacing $\Delta~r=0.6$~fm and a time step $\Delta~t=5\times10^{-25}$~s.
The size of the half-box where the single particle wave functions are evolved is $80\times80\times40$ in mesh size unit $\Delta~r$, unless otherwise specified.

\section{Results}
\label{sec:results}

Let us now investigate the coupling between isovector dipole and isoscalar quadrupole vibrations 
in tin isotopes in the framework of the theoretical model presented in Sec.~\ref{sec:model}
where $|\nu\>\equiv |D\>$ and  $|\mu\>\equiv |Q\>$ denote a GDR and a GQR phonon respectively.
The isovector dipole moment is defined as 
\oeq
\oQ_D = \frac{NZ}{A} (\oZ_n-\oZ_p)
\ceq
where $\oZ_{n}$ (resp. $\oZ_{p}$) measures the neutron (resp. proton) average position on the $z$ axis.
The isoscalar quadrupole moment reads
\oeq
\oQ_Q = \sqrt{\frac{5}{16\pi}} \sum_{i=1}^A (2\oz_i^2-\ox_i^2-\oy_i^2).
\label{eq:Qq}
\ceq
Their expectation value evolutions are computed using the
{\textsc{tdhf3d}} code after different initial conditions as described below.

\subsection{Nonlinear quadrupole motion induced by a dipole boost}
\label{subsec:TDHFquadratic}

\begin{figure}
\includegraphics[width=8cm]{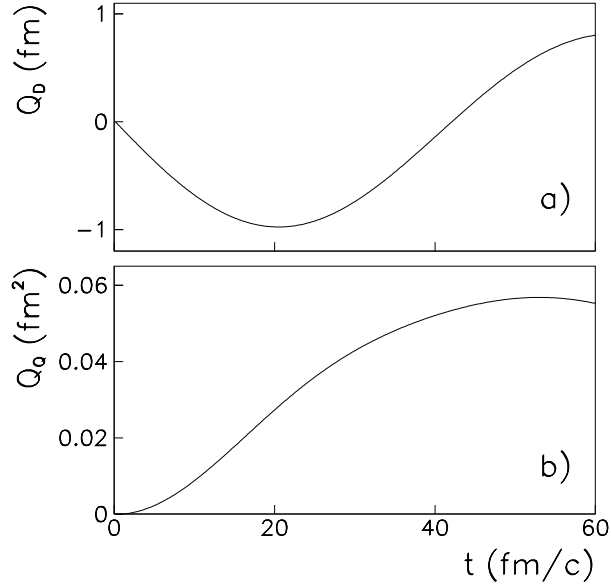}
\caption{Time evolution of the dipole (a) and quadrupole (b) moments in $^{132}$Sn 
after a dipole boost with an intensity $k_D=0.01$~fm$^{-1}$.
\label{fig:time}}
\end{figure}

We first investigate the quadratic response presented in Sec.~\ref{subsubsec:quadratic} in the $^{132}$Sn nucleus.
Figure~\ref{fig:time}(a) shows the early time evolution of the dipole moment
after a dipole boost according to Eq.~(\ref{eq:boost}) in the small-amplitude regime.
The dipole moment follows a $-\sin$ function as indicated by Eq.~(\ref{eq:linear}).
Extracting the frequency from the first minimum of $D(t)$ leads to a GDR energy of $\omega_D\approx15.2$~MeV. This value is slightly lower than the maximum of the experimental GDR peak energy $E_{max}=16.1(7)$~MeV~\cite{adr05}. 
The same analysis in $^{120}$Sn, in which almost all the dipole strength 
is located around the GDR energy, gives a value of $\omega_D^{(^{120}Sn)}\approx15.3$~MeV which is in good agreement with experimental data where a peak energy of $E_{GDR}^{exp.}=15.4$~MeV 
has been obtained~\cite{ber75}. 
Note that, the extraction method of $\omega_\nu$ from the first extremum of $Q_\nu(t)$ is, 
in first approximation, comparable to the ratio of the second over the first energy weighted 
moments of the strength function~$m_2/m_1$~\cite{sim03}.

We see in figure~\ref{fig:time}(b) that an oscillation of the quadrupole moment is induced by the dipole boost.
According to the theoretical model presented in Sec.~\ref{sec:model}, 
this is a manifestation of the residual interaction of Eq.~(\ref{eq:H}) coupling the dipole and quadrupole vibrations.  
In particular, $Q_Q(t)$ starts in phase quadrature with $Q_D(t)$ and oscillates with a smaller frequency.
These observations are in qualitative agreement with the quadratic response in Eq.~(\ref{eq:quadratic}).

\begin{figure}
\includegraphics[width=8cm]{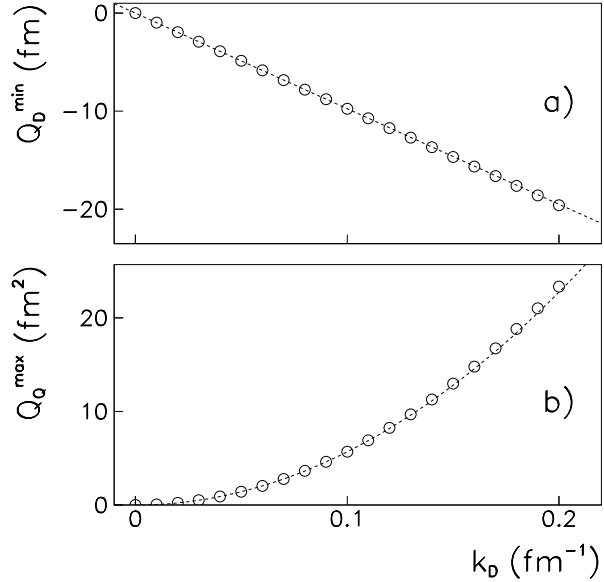}
\caption{Circles: First minimum and maximum of the dipole (a) and quadrupole (b) moment evolution, respectively, in $^{132}$Sn 
as a function of the dipole boost intensity~$k_D$. 
Dashed lines: linear and quadratic extrapolations at $k_D\rightarrow0$ of the dipole (a) and quadrupole (b) amplitudes, respectively.
\label{fig:quadratic}}
\end{figure}

To get a deeper insight into this coupling, we have computed the TDHF response for
several dipole boost velocities~$k_D$.  The first extrema of the dipole and quadrupole 
moments are reported in figure~\ref{fig:quadratic}(a) and (b) respectively.
Whereas the dipole amplitude is indeed linear in $k_D$
as expected from equation~(\ref{eq:linear}), indicating that 
these calculations are performed in the small-amplitude regime,
the induced quadrupole motion is quadratic in $k_D$, 
in agreement with Eq.~(\ref{eq:quadratic}).

To obtain a quantitative estimate of the coupling, we first extract 
the transition amplitude from a linear extrapolation of $Q_D^{min}$ 
at $k_D\rightarrow0$ in Fig.~\ref{fig:quadratic}(a). 
According to Eq.~(\ref{eq:linear}), we get $q_D\approx6.98$~fm.
The same analysis with a quadrupole boost in the linear regime 
gives a transition amplitude $q_Q\approx61.4$~fm$^2$ and a GQR
energy of $\omega_Q\approx13.0$~MeV. Note that the same analysis in $^{120}$Sn gives a GQR energy of $\omega_Q^{(^{120}Sn)}\approx13.3$~MeV, in excellent agreement with 
the experimental value $E_{GQR}^{exp.}=13.24\pm0.13$~MeV~\cite{sha88}.
These quantities, together with a quadratic extrapolation 
of the quadrupole maximum at $k_D\rightarrow0$ in Fig.~\ref{fig:quadratic}(b),
give, according to Eq.~(\ref{eq:quadratic}), a matrix element of the residual
interaction~$v_Q^{(1)}\approx-0.61$~MeV.

\subsection{Dipole motion in a nucleus with a static quadrupole constraint}

The formalism developed in Sec.~\ref{subsubsec:deformed}, 
where the linear response is investigated in an external potential,
cannot be directly applied to study the coupling between the dipole and quadrupole modes.  
The reason is that the external potential $-\lambda\oQ_Q$ with the definition
of Eq.~(\ref{eq:Qq}) is not bound from below
and its use in constrained HF calculations would lead to unphysical results.
It is then necessary to consider another external potential such as 
\oeq
\lambda (\oQ_Q+\kappa_\lambda\oQ_M)
\label{eq:constraint}
\ceq
where 
\oeq
\oQ_M=\frac{1}{\sqrt{4\pi}} \sum_{i=1}^A \ovr_i^2
\ceq 
is the monopole moment and $\kappa_\lambda=\sqrt{5}/2$ if $\lambda\geq0$ and $-\sqrt{5}$ if $\lambda<0$.
The expression~(\ref{eq:constraint}) then reads
\oeq
3\sqrt{\frac{5}{16\pi}}\lambda\sum_{i=1}^A\left\{
\begin{array}{cl}
\oz_i^2 & \mbox{if~}\lambda\geq0, \\
-\ox_i^2-\oy_i^2 & \mbox{if~} \lambda<0.
\end{array}
\right.
\ceq

\begin{figure}
\includegraphics[width=8cm]{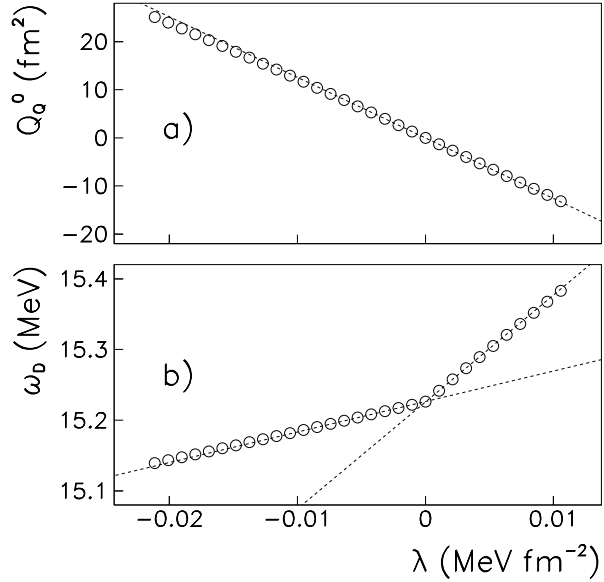}
\caption{(a) Static quadrupole moment from HF calculation (circles) under a quadrupole+monopole constraint (see text) as a function of the Lagrange parameter~$\lambda$ in~$^{132}$Sn.
(b) TDHF energy of the GDR (circles) from the first minimum of the dipole moment after a dipole boost along the deformation axis
with an intensity $k_D=0.01$~fm$^{-1}$. 
Dashed lines : linear extrapolations at $\lambda\rightarrow0^{\pm}$ of the quadrupole moment (a) and GDR energy (b).
\label{fig:deformed}}
\end{figure}

Such an external field allows one to explore all quadrupole deformations from oblate ($\lambda>0$) to prolate ($\lambda<0$) shapes 
as shown in figure~\ref{fig:deformed}(a) where the ground state quadrupole deformation $Q_Q^0$ of the constrained HF solution 
is plotted as a function of the Lagrange parameter $\lambda$.
The quadrupole deformation is clearly linear in this perturbative regime 
and its slope at the origin is 
$\left.\frac{\partial Q_Q^0}{\partial \lambda}\right|_{\lambda=0}\approx-1260.4$~fm$^4$~MeV$^{-1}$.

As discussed in Sec.~\ref{subsubsec:deformed}, such a static deformation is expected to change the 
dipole frequency as compared to that of the GDR excited on the spherical ground state.
In fact, the frequency of a dipole oscillation along the main quadrupole axis decreases (resp. increases)  
with a prolate (resp. oblate) deformation.
This is indeed what we observe in figure~\ref{fig:deformed}(b) where the energy of the GDR is plotted 
as a function of $\lambda$.
Note that, according to Eq.~(\ref{eq:slope}), this is consistent 
with the negative sign of the ratio $v_Q/q_Q$ obtained in Sec.~\ref{subsec:TDHFquadratic}. 

We also observe in Fig.~\ref{fig:deformed}(b) that 
the evolution of this energy is linear both for $\lambda>0$ and $\lambda<0$, 
but the slopes are different in these two regimes.
This is attributed to the presence of the monopole moment in the constraint~(\ref{eq:constraint}).
Indeed, the monopole vibration is also coupled to the dipole mode 
by a matrix element $v_M$ of the residual interaction~\cite{fal03,sim03}. 
According to Eq.~(\ref{eq:slope}), the dipole energy is expected to be modified as
\oeq
\omega_D(\lambda) \approx \omega_D(0) 
+ \lambda\frac{v_Q}{q_Q}\(\frac{\partial Q_Q^0}{\partial \lambda}\)_{\lambda=0}
+ \lambda\kappa_\lambda\frac{v_M}{q_M}\(\frac{\partial Q_M^0}{\partial \lambda}\)_{\lambda=0}.
\label{eq:def}
\ceq
A compression of the nucleus increases the dipole frequency, which implies $v_M/q_M<0$.
Because $\lambda\kappa_\lambda\ge0$ for all $\lambda$, the monopole and quadrupole moments
have an opposite effect on $\omega_D$ for $\lambda<0$ and act in the same direction for $\lambda>0$.
This is indeed what we observe in Fig.~\ref{fig:deformed}(b) where the effect of the constraint almost cancels
on the prolate side, whereas it strongly increases the GDR energy in the oblate one.

Finally, starting from Eq.~(\ref{eq:def}), it is possible to isolate the coupling matrix element 
between the dipole and quadrupole modes
\oeq
v_Q=\(\frac{\partial Q_Q^0}{\partial \lambda}\)_{\lambda=0}^{-1} \frac{q_Q}{3} 
\[ \( \frac{\partial \omega_D}{\partial \lambda}\)_{\lambda\rightarrow0^-} + 2
 \(\frac{\partial \omega_D}{\partial \lambda}\)_{\lambda\rightarrow0^+} \].
\ceq
Using the data extracted from Fig.~\ref{fig:deformed} 
and the value of $q_Q$ obtained in Sec.~\ref{subsec:TDHFquadratic},
we get $v_Q^{(2)}\approx-0.56$~MeV. 
This result is in reasonable agreement with the one obtained with the quadratic response.

\subsection{Response to a dipole+quadrupole boost}

A third manifestation of the coupling between dipole and quadrupole motions occurs when 
both a dipole and a quadrupole boost are performed at initial time. 
We showed in Sec.~\ref{subsubsec:double_boost} that, in such a case, 
the dipole motion is affected by the quadrupole vibration.
Such effect is not present in the linear response theory because the modifications are proportional to $k_Dk_Q$.
As can be seen in Eq.~(\ref{eq:linear}), there is no other quadratic term 
because the next-order terms affecting the dipole motion are in~$k_D^3$ and~$k_Q^3$.

The basic tool to study the effect of the coupling on the dipole motion 
is the coupling response function defined in Eq.~(\ref{eq:coupling}).
In principle, its calculation implies to follow the dipole moment over an infinite time. 
However, we use a filtering procedure to avoid numerical artifacts coming from 
the interaction of the nucleus with reflected nucleon wave functions 
because of the hard box boundary conditions~\cite{rei06}.
We perform the calculations over 3000 iterations in time (450~fm/$c$).
The dipole moment is multiplied by a filtering function
$\exp\[-\frac{1}{2}\(\frac{t}{\tau}\)^2\]$ with $\tau=100$~fm/$c$~\cite{mar05}. 
This procedure induces an additional width of $\tau^{-1}\approx 2$~MeV.
According to Eq.~(\ref{eq:resp_func}), this additional width is sufficiently small 
for the present discussion because the modes in the coupling response function are
located at $\omega_D-\omega_Q\approx2.2$, $\omega_D\approx15.2$, and $\omega_D+\omega_Q\approx28.2$~MeV. 
However, the low-energy part of the spectrum, i.e., in the region of the $\omega_D-\omega_Q$ peak,
is dependent on the choice of the filtering function within these numerical conditions. 
We checked with other filtering functions, e.g., a cosine instead of a Gaussian function, 
to confirm that the higher part of the spectrum (above $\approx~10$~MeV) is not affected.
In addition, the filtering function does not change the fact that the total integral of the 
coupling response function vanishes (see Sec.~\ref{subsubsec:double_boost}). 
The latter was found to be a solid numerical property of this function.
Finally, we checked the convergence of the results presented in this section
by comparing with calculations performed in a bigger box of $120\times120\times60$ in mesh size unit $\Delta~r$.

\begin{figure}
\includegraphics[width=8cm]{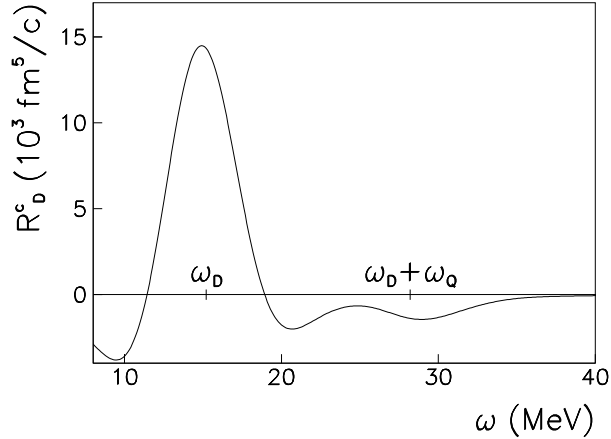}
\caption{Coupling response function of the dipole moment after a dipole+quadrupole boost with intensities $k_D=0.01$~fm$^{-1}$ and $k_Q=0.001$~fm$^{-2}$ respectively.
\label{fig:tf}}
\end{figure}

Figure~\ref{fig:tf} shows the coupling response function  
for the dipole motion following a quadrupole+dipole boost.
We checked that, in the small-amplitude limit, 
the coupling response function is indeed independent of $k_{Q}$ and $k_D$.
As expected from Eq.~(\ref{eq:resp_func}), two peaks are present in this energy range
at $\omega_D$ and $\omega_D+\omega_Q$ with opposite signs.
Moreover, 
the integral of the positive peak at $\omega_D$ is directly related to the coupling as
\oeq
v_Q=-\frac{\omega_Q}{2 q_D^2q_Q}\int_{R^c_D>0}\stb \d \omega \stf R^c_D(\omega).
\ceq
With the coefficients calculated in Sec.~\ref{subsec:TDHFquadratic},
we obtain a coupling $v^{(3)}\approx-0.68$~MeV of the same order of magnitude than with the two previous methods.

Let us finally note that in the case of more complicated vibrations, e.g., the oscillations exhibit several frequencies, 
the coupling response function can be used for a more detailed investigation of the coupling.
Indeed, it allows an analysis of the coupling effect at each energy
whereas the two previous methods give only access to a weighted sum 
of the matrix elements of the residual interaction associated to each excited mode~\cite{sim03}. 

\subsection{Evolution of the coupling with isospin and mass}

We now repeat the study of the linear quadrupole motion induced by a dipole boost,
described in Sec.~\ref{subsec:TDHFquadratic}, to the tin isotopic chain.
The choice of this method to investigate more systematically 
the coupling between dipole and quadrupole vibrations 
is motivated by its rather low computational time as compared to the two other methods.
Our goal is to understand the evolution of $v_Q$ as a function of the isospin.
To avoid any ambiguity coming from possible static deformation in the ground states,
we focus on some of the tin isotopes that are spherical at the HF level:
$^{100,106,114,120,132,140}$Sn. 
These isotopes allow for an investigation of the coupling 
from the proton-rich to the neutron-rich side.

\begin{figure}
\includegraphics[width=8cm]{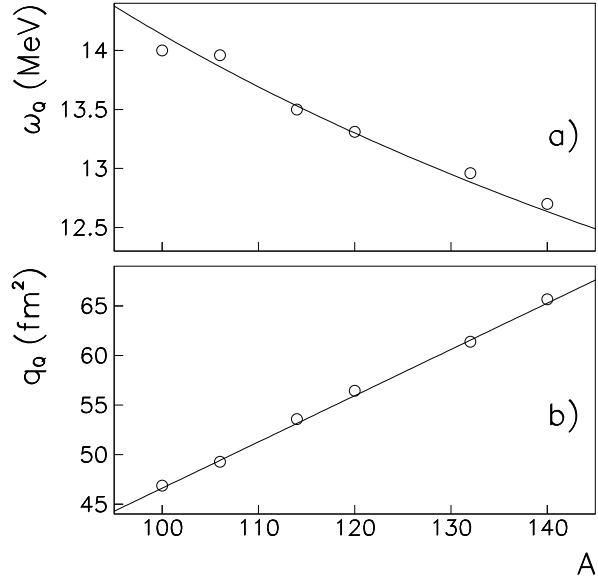}
\caption{Evolution of (a) the GQR energy and (b) the transition amplitude 
as a function of the number of nucleons in tin isotopes from the TDHF linear response (circles).
The line in panel (a) represents a $A^{-1/3}$ fit of the TDHF results, 
whereas the line in panel (b) is obtained from considerations on the GQR energy weighted sum rule (see text).
}
\label{fig:quadrupole_Sn}
\end{figure}

Let us first investigate the linear response to a quadrupole boost (Eq.~(\ref{eq:boost}))
 to compute the energies~$\omega_Q$ and transition amplitudes~$q_Q$
from the first minimum of the quadrupole moment (see Eq.~(\ref{eq:linear})).
These quantities are plotted in Fig.~\ref{fig:quadrupole_Sn} as a function of the number
of nucleons. The GQR energy is known to be proportional to $A^{-1/3}$~\cite{har01}.
This is compatible with the TDHF results that are fitted by~$\omega_Q\approx~65.5A^{-1/3}$~MeV.
The evolution of the transition amplitude with~$A$ can be obtained from the 
energy weighted sum rule (EWSR) for quadrupole vibrations which reads~\cite{rin80}
\oeqn
S_Q^1 &=& \sum_{\al} (E_\al-E_0) |\<\al|\oQ_Q|0\>|^2 \nonumber \\
&=& \frac{\hb^2}{m} \frac{5}{4\pi} A \<\ovr^2\> \nonumber \\
&\approx&  14.3 A^{5/3}{\mbox{ MeV.fm$^4$}}
\label{eq:SQ1}
\ceqn
where $\{\kal\}$ is an eigenbasis of $\oH$. 
In the last line of Eq.~\ref{eq:SQ1}, we used the approximation of a constant density 
and a sharp surface which gives $\<\ovr^L\>=\frac{3}{L+3}R^L$ with $R\approx~1.2A^{1/3}$~fm.
If all the strength is located at the GQR energy, 
which is a somewhat good approximation for heavy nuclei~\cite{har01},
then the EWSR reduces to
\oeq
S_Q^1 = \omega_Q q_Q^2\approx 65.5A^{-1/3}q_Q^2.
\label{eq:SQ2}
\ceq
Equations~(\ref{eq:SQ1}) and~(\ref{eq:SQ2}) then lead to 
\oeq
q_Q\approx 0.466A{\mbox{ fm$^2$}}.
\label{eq:q_A}
\ceq
This linear dependence is plotted in Fig.~\ref{fig:quadrupole_Sn}(b) and reproduces well the TDHF results.

\begin{figure}
\includegraphics[width=8cm]{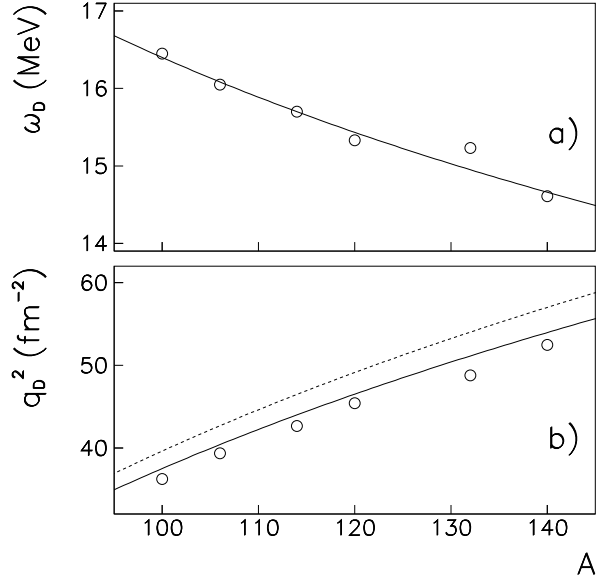}
\caption{Evolution of (a) the GDR energy and (b) the transition probability 
as a function of the number of nucleons in tin isotopes from TDHF linear response (circles).
The line in panel (a) represents a $A^{-1/3}$ fit of the TDHF results.
The lines in panel (b) is obtained from considerations on the GDR energy weighted sum rule (see text)
with an enhancement factor of the Thomas-Reiche-Kuhn sum rule $\kappa=0.25$ (dashed line)
and $\kappa=0.183$ (solid line).
}
\label{fig:dipole_Sn}
\end{figure}

Let us now consider a dipole boost on these nuclei with a boost velocity $k_D=0.01$~fm$^{-1}$. 
This value is small enough to generate a linear response of the dipole moment
and a quadratic response of the induced quadrupole vibration in all considered isotopes.
The GDR energy~$\omega_D$  is shown as a function of the number of nucleons in Fig.~\ref{fig:dipole_Sn}(a).
It is compatible with the $A^{-1/3}$ dependence expected in heavy nuclei~\cite{har01}.
A fit of the TDHF results gives
\oeq
\omega_D\approx 76A^{-1/3}{\mbox{~MeV.}}
\label{eq:omega_D}
\ceq
Similarly to the quadrupole case, the dependence of the transition probability $q_D^2$ 
can be obtained from the dipole EWSR 
\oeq
S_D^1 = \frac{\hbar^2}{2m}(1+\kappa)\frac{NZ}{A}
\label{eq:SD}
\ceq
where $\kappa$ is the enhancement factor of the Thomas-Reiche-Kuhn (TKR) sum rule.
Assuming all the strength in the GDR, i.e., $q_D^2=S_D^1/\omega_D$, we get from Eqs.~(\ref{eq:omega_D}) and~(\ref{eq:SD})
\oeq
q_D^2\approx \frac{\hbar^2}{2m}(1+\kappa) \frac{NZ}{76A^{2/3}}
\label{eq:q_D}
\ceq
In nuclear matter, the enhancement factor of the TKR sum rule is $\kappa=0.25$ with the SLy4 parametrization~\cite{cha98}.
This value clearly overestimates the transition probabilities (see dashed line in Fig.~\ref{fig:dipole_Sn}(b)), 
though the qualitative trend is in good agreement with the TDHF results. 
It is possible to compute  $\kappa$ in finite nuclei using~\cite{cha97}
\oeq
\kappa = \frac{m}{4\hbar^2}\,\frac{A}{NZ}\[t_1(2+x_1)+t_2(2+x_2)\]\int \d r^3 \rho_n(\vr) \rho_p(\vr)
\label{eq:kappa}
\ceq
where the $t_i$ and $x_i$ are the usual Skyrme parameters.
In the considered tin isotopes, $\kappa$ is almost constant within the range $0.181-0.186$ 
with no particular isospin or mass dependence. 
This leads to a better agreement with the transition probabilities obtained with TDHF 
(see solid line in Fig.~\ref{fig:dipole_Sn}(b)), though a slight overestimation remains.
The latter could be attributed to the fragmentation of the isovector dipole response.
In this case, the dipole response reads
\oeq 
Q_D(t) = -2k_D \sum_i q_{D_i}^2 \sin[(\omega_D+\delta \omega_i)t].
\ceq
In our calculations, the GDR properties ($\omega_D$ and $q_D$) are extracted from the first minimum of the dipole response, which obeys to 
\oeq
\frac{Q_D^{min}}{-2k_D}= \sum_iq_{D_i}^2\cos\(\frac{\pi}{2}\frac{\delta \omega_i}{\omega_D}\)\equiv q_D^2 \leq \sum_i q_{D_i}^2.
\ceq
The last inequality implies that the TKR sum rule allows one only to compute the upper limit of $q_D^2$ in our model.

Finally, we investigate the coupling between the quadrupole and dipole vibrations
from the quadratic response. 
We have shown in Sec.~\ref{subsubsec:quadratic} that, in the presence of a nonzero matrix element $v_Q$ 
of the residual interaction coupling the state $|D\>$ to the state $|DQ\>$, a dipole boost
is expected to generate an oscillation of the quadrupole moment.
Before studying the evolution of~$v_Q$ along the tin isotopic chain, 
it is mandatory to get a deeper insight into the mechanism 
responsible for this induced quadrupole excitation.

In a macroscopic approach, the isovector GDR is interpreted by a combination
of the Steinwedel-Jensen model in which the total density is kept unchanged~\cite{ste50} 
and the Goldhaber-Teller model where proton and neutron fluids are incompressible~\cite{gol48}.
It is obvious that the Steinwedel-Jensen model does not affect 
 the quadrupole moment because any modification of the density of one isospin specie 
is exactly compensated by the other in every point of space. 
In the Goldhaber-Teller model, however, a displacement of the proton and neutron spheres 
in the opposite direction is considered. 
It produces a dipole moment $Q_D=\frac{NZ}{A}X$ where $X$ is the distance between their centers. 
This displacement also induces a prolate shape with a quadrupole moment quadratic in $X$.
Indeed, assuming a displacement of a proton (resp. neutron) homogeneous sphere of density $Z\rho_0/A$ 
(resp. $N\rho_0/A$) by $X_p=-XN/A$ (resp. $X_n=XZ/A$) produces a quadrupole moment
\oeq
Q_Q\propto~ZX_p^2+NX_n^2=\frac{NZ}{A}{X^2}.
\ceq
Using Eqs.(\ref{eq:linear}) and~(\ref{eq:q_D}), one gets 
$Q_Q\propto~NZ/A^{1/3}$.

\begin{figure}
\includegraphics[width=8cm]{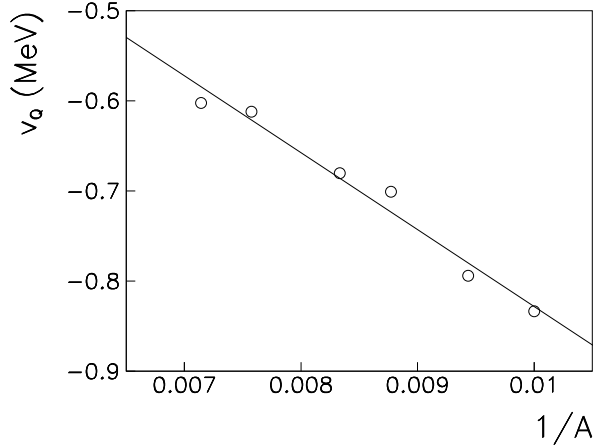}
\caption{Evolution of the coupling with mass number.
The matrix element of the residual interaction is plotted as a function of $1/A$.
The line shows a linear fit of the TDHF results.}
\label{fig:v}
\end{figure}

Finally, together with Eqs.~(\ref{eq:quadratic}), (\ref{eq:q_A}) and (\ref{eq:q_D}), 
the evolution of the coupling simply reads $v_Q\propto~1/A$.
This is, indeed, in agreement with the TDHF results shown in Fig.~\ref{fig:v}.
It is interesting to note that, in this simple approach, 
the coupling does not depend on the isospin of the nuclei, 
but only on their total number of nucleons.
In fact, the decrease of the absolute strength of the coupling with the number of nucleons
is attributed to the fact that these couplings are mediated by the surface~\cite{sim03}.
One then expects less anharmonicities in heavy nuclei.

\section{conclusions}
\label{sec:conclusion}

We have shown that the residual interaction is responsible for anharmonicities in nuclear vibrations
using three different analyses of time evolutions of multipole moments.
We investigated the coupling between one- and two-phonon states using a 3-dimensional TDHF code
with a full Skyrme energy density functional.
In particular, the excitation of a GDR couples to a GQR built on top of it, 
inducing a quadratic response of the quadrupole moment. 
The same coupling is responsible for the change of the GDR energy in static deformed states. 
The latter could be investigated using deformed RPA codes.
As a consequence, the dipole frequency is modulated in case of dynamical deformation, 
e.g., induced by a quadrupole boost. This last property, associated with a Fourier analysis, 
might be used to investigate couplings when more than one mode is excited with the same quantum numbers. 
We finally investigated these couplings with the quadratic response in several spherical tin isotopes.
As a result, no dependence with isospin was found whereas an overall decrease of the coupling is obtained
with increasing mass, showing that the couplings are mediated by the surface.
These observations are interpreted within the Goldhaber-Teller macroscopic model.
These results indicate that no anharmonicity enhancement is expected in the standard giant resonances even for very exotic nuclei. However, couplings with exotic modes such as the pygmy dipole resonance should be investigated with the present method. The role of pairing and static deformation should be considered as well. 
\begin{acknowledgements} 
We thank P.~Bonche for providing his TDHF code.
We are also grateful to B.~Avez and D.~Lacroix for discussions and a careful reading of the article. 
Useful discussions with K.~Bennaceur are also acknowledged. The calculations were performed at the 
Centre de Calcul Recherche et Technologie of the Commissariat \`a l'\'Energie Atomique.
\end{acknowledgements}


\end{document}